\documentclass[12pt,epsf]{article}

\usepackage{graphicx}
\usepackage{mathrsfs}
\usepackage{amssymb}
\usepackage{amsmath}

\newcommand{\be}{\begin{equation}}

\newcommand{\ee}{\end{equation}}

\newcommand{\ba}{\begin{array}}

\newcommand{\ea}{\end{array}}

\begin{document}
\begin{titlepage}
\vspace{.5in}
\begin{flushright}
CQUeST-2009-0297
\end{flushright}
\vspace{0.5cm}

\begin{center}
{\Large\bf Instanton solutions mediating tunneling between the degenerate vacua in curved space}\\
\vspace{.4in}

  {$\rm{Bum-Hoon \,\, Lee}^{\dag}$}\footnote{\it email:bhl@sogang.ac.kr}\,\,
  {$\rm{Chul \,\, H. \,\, Lee}^{\S}$}\footnote{\it email:chulhoon@hanyang.ac.kr}\,\,
  {$\rm{Wonwoo \,\, Lee}^{\dag}$}\footnote{\it email:warrior@sogang.ac.kr}\,\,
  {$\rm{Changheon \,\, Oh}^{\S\ddag}$}\footnote{\it email:och0423@hanyang.ac.kr}\\

  {\small \dag \it Department of Physics and BK21 Division, and Center for Quantum Spacetime, Sogang University, Seoul 121-742, Korea}\\
  {\small \S \it Department of Physics, Hanyang University, Seoul 133-791, Korea}\\
  {\small \ddag \it Principal Researcher Center, Technovation Partners, Seoul 135-824, Korea}\\

\vspace{.5in}
\end{center}
\begin{center}
{\large\bf Abstract}
\end{center}
\begin{center}
\begin{minipage}{4.75in}

{\small \,\,\,\,We investigate the instanton solution between the
degenerate vacua in curved space. We show that there exist
$O(4)$-symmetric solutions not only in de Sitter but also in both
flat and anti-de Sitter space. The geometry of the new type of solutions is finite and preserves the $Z_2$ symmetry. The nontrivial solution corresponding to the tunneling is possible only if gravity is taken into account. The numerical solutions as well as the
analytic computations using the thin-wall approximation are
presented. We expect that these solutions do not have any negative mode as in the instanton solution.}


\end{minipage}
\end{center}
\end{titlepage}

\newpage
\section{Introduction \label{sec1}}

The instanton solution with $O(4)$ symmetry exists between the degenerate vacua in de Sitter(dS) space. This can be understood in the particle analogy picture due to the change of the damping term into the accelerating term. The numerical solution of the scalar field $\Phi$ was obtained in Ref.\ \cite{hw02}. The analytic computation and meanings of this solution were further studied in Ref.\ \cite{bl03}. Can we also obtain the instanton solution between the degenerate vacua in both flat and anti-de Sitter(AdS) space? The aim of this paper is to obtain solutions in these backgrounds. We study boundary conditions needed for these solutions not only in dS but also in both flat and AdS space. These boundary conditions respect the $Z_2$ symmetry, different from those studied in bounce solutions \cite{voloshin, col002, bnu02, par02}. We will show that there exist a new type of solutions giving rise to the finite geometry with the  $Z_2$ symmetry.

An instanton solution is a solution to the equation of motion of the classical field theory in Euclidean space satisfying appropriate boundary conditions. Yang-Mills instantons give finite action, and have been extensively studied in gauge theories \cite{bpst} as well as in string theory \cite{ggp} (for a review, see Ref.\ \cite{vni}). In gravitational theory, there are several kinds of instantons \cite{ghh}. A recent summary is in Ref.\ \cite{page} and references therein. Some of those have finite action, and others an infinite one \cite{hb0}. There is also an issue on the sign of the action for
the solutions describing quantum creation of the inflationary universe \cite{vhl}. Some of the authors in Ref.\ \cite{vhl} considered the counterclockwise Euclidean rotation rather than the clockwise rotation as the analytic continuation in the complex $t$ plane to suppress the probability with negative action. This is related to the fact that the Euclidean Einstein action is not positive definite, which is known as the conformal factor problem in Euclidean quantum gravity \cite{ghp000}. However, this issue is not central to the problem and therefore we will not discuss the issue any more in the present paper.

Quantum mechanically, the tunneling process in a symmetric double-well potential is described by the instanton solution. This tunneling shifts the ground state energy of the classical vacuum due to the presence of an additional potential well lifting the classical degeneracy. The symmetric ground state wave function describes that there is a higher probability of finding it somewhere between the two classical vacua. In the semiclassical approximation, the action is dominated by the instanton configuration. The instanton solution can be interpreted as a particle motion in the inverted potential starting from one vacuum state at the minus infinite Euclidean time and reaching the other one at the plus infinite time. We may consider the multi-instanton solutions as describing the tunneling back and forth between the two vacua. For field theoretical solutions with $O(4)$ symmetry, the equation of motion has an additional term, which can be interpreted as a friction term in the inverted potential. The solution for the potential with degenerate minima is known for the dS case. However, there are no studies of the solution in flat or AdS space.

The tunneling solution in the field theory with an asymmetric potential is called a bounce solution. The solution is related to
the nucleation of a true (false) vacuum bubble describing the decay of a metastable state. There have been many studies on the decay of a metastable vacuum. The process was first investigated in Ref.\ \cite{voloshin} and developed in both flat \cite{col002} and curved spacetime \cite{bnu02, par02}. Another type of transition describing the scalar field jumping simultaneously at the top of the potential barrier was investigated by Hawking and Moss \cite{hawking} and later in Ref.\ \cite{jst01}. Further recent studies can be found in Refs.\ \cite{marvel01, bj05}.

The mechanism for the nucleation of a false vacuum bubble within
the true vacuum background has also been studied. The nucleation
process of the large false vacuum bubble in dS space was originally
obtained in Ref.\ \cite{kw04}, and the case with a global monopole in Ref.\ \cite{yms}. The small false vacuum bubble in
the Einstein theory of gravity with a nonminimally coupled scalar
field was explored using a mechanical analogy in Ref.\
\cite{wl01}, and the case with the Gauss-Bonnet term in Ref.\ \cite{koh01}. The nucleation rate as well as classification
depending on the size of the false vacuum bubble in dS space was
obtained in Ref.\ \cite{bl03} in the Einstein theory of gravity. The classification of true and false vacuum bubbles in the dS background space depends on the value of the cosmological constant. Large bubbles have large values of the cosmological constant, while small bubbles have small values of the cosmological constant.
These processes may provide an alternative paradigm in the string theory landscape \cite{landscape01} or eternal inflation \cite{eternal01}.

The natural question is on the instanton solution between the
degenerate vacua and its relation to the bubble. The clue of these solutions could have been seen in Refs.\ \cite{hw02, bl03, bj05}. The instanton solution in the potential with degenerate vacua in de Sitter background was obtained numerically \cite{hw02}. It was also shown that the solution can be analyzed as the limiting case of the large true vacuum bubble or large false vacuum bubble \cite{bl03}. In this paper, we will show that there also exist solutions between the degenerate vacua in both flat and AdS backgrounds. These solutions can not be obtained as the limiting cases of the previously known vacuum bubble solutions.

In flat spacetime the bounce solution has exactly one negative mode \cite{nem}. When gravity is taken into account, it is a more involved problem \cite{nemode}. On the other hand, our solutions describe quantum mechanical mixing between the degenerate vacua and we expect that the solutions do not have any negative mode.

The time evolution of the solution after its materialization can be studied by the analytic continuation to Lorentzian spacetime. We will study the dynamics of the wall following the introductory work of Coleman \cite{col002}. The wall has a trajectory of the hyperboloid in Minkowski spacetime. The method in curved spacetime was also studied \cite{chw02}. The metric junction condition \cite{isr01} can also be employed for the evolution of the wall (for recent works, see Ref.\ \cite{dyb} and references therein). The evolution is simply classified into two types from observer's point of view on the wall: One is shrinking while the other is expanding. For the type of the
shrinking wall, there are two cases. One is related to the
creation of a child universe. The other is related to the black
hole formation or nothing. For the type of the expanding wall,
there are also two cases. One is the bubble eating up the original
background. The other is the case with the expanding inside region
as well as the outside region, which will be related to the
present work.

The outline of this paper is as follows: In Sec.\ 2 we investigate
the instanton solution with $O(4)$ symmetry between the degenerate
vacua in not only de Sitter space but also flat and AdS space. We
discuss the new type of boundary conditions with $Z_2$ symmetry. Under these boundary conditions, we explore the possibility for the existence of the instanton solution qualitatively. From this perspective, one of the central questions that we intend to is to obtain the condition for the existence of the solution. The condition will depend on the local maximum value of the potential. In Sec.\ 3 we employ the metric junction condition to get the dynamics of the wall. In Sec.\ 4 we summarize and discuss our results.

\section{The instanton solution \label{sec2}}

We consider the following action
\begin{equation}
S= \int_{\mathcal M} \sqrt{-g} d^4 x \left[ \frac{R}{2\kappa}
-\frac{1}{2}{\nabla^\alpha}\Phi {\nabla_\alpha}\Phi
-U(\Phi)\right] + \oint_{\partial \mathcal M} \sqrt{-h} d^3 x
\frac{K}{\kappa}, \label{f-action}
\end{equation}
where $\kappa \equiv 8\pi G$, $g\equiv det g_{\mu\nu}$, and the
second term on the right-hand side is the boundary term \cite{ygh}.

The vacuum-to-vacuum  transition amplitude can be semiclassically represented as $A e^{-\Delta S}$, where the exponent $\Delta S$ is the difference between the Euclidean action corresponding to an instanton solution and the background action itself. This exponent is also related to the splitting of the energy levels for the symmetric double-well potential. The prefactor $A$ comes from the first order quantum correction.

If we consider the asymmetric double-well potential, the
metastable state will tunnel into the ground state through the
bubble nucleation process. The action corresponding to the
tunneling process will then have an imaginary part. The decay rate of the background vacuum is described by a bounce solution
\cite{voloshin, col002, bnu02, nem, prefactor}.

Our main concern in this paper is the case of the symmetric double-well potential. The
scalar potential $U(\Phi)$ in Eq.\ (\ref{f-action}) has two
degenerate minima
\begin{equation}
U(\Phi) = \frac{\lambda}{8} \left(\Phi^2-
\frac{\mu^2}{\lambda}\right)^2 + U_o. \label{poten}
\end{equation}
The space will be dS, flat, or AdS depending on whether $U_o > 0$,
$U_o = 0$, or $U_o < 0$. We want to obtain instanton solutions describing the tunneling between the degenerate vacua in the present work.

We consider $O(4)$-symmetric ansatz for both the geometry and the scalar field
\begin{equation}
ds^2 = d\eta^2 + \rho^2(\eta)[d\chi^2 +
\sin^2\chi(d\theta^2+\sin^2\theta d\phi^2)].  \label{gemetric}
\end{equation}
Then, $\Phi$ and $\rho$ depend only on $\eta$ and the Euclidean field equations for them have the
form
\begin{equation}
\Phi'' + \frac{3\rho'}{\rho}\Phi'=\frac{dU}{d\Phi} \,\,\,\, {\rm
and} \,\,\,\, \rho'' = - \frac{\kappa}{3}\rho (\Phi'^2 +U),
\label{ephi}
\end{equation}
respectively and the Hamiltonian constraint is given by
\begin{equation}
\rho'^2 - 1 -
\frac{\kappa\rho^2}{3}\left(\frac{1}{2}\Phi'^{2}-U\right) = 0 .
\label{erho}
\end{equation}

We now have to impose the boundary conditions to solve Eqs.\ (\ref{ephi}) and (\ref{erho}). In this work, we have to consider $4$ conditions for two equations of $2$nd order. They can be the values of the fields $\Phi$ and $\rho$, or their derivatives $\Phi'$ and $\rho'$ at either $\eta=0$ or at $\eta=\eta_{max}$. For example, in flat spacetime without gravity where $\rho$ is identified with $\eta$, $\Phi$ at $\eta_{max}$, and $\Phi'$ at $\eta=0$ was imposed as the boundary conditions for $\Phi$ in Ref.\ \cite{col002}. The condition $\Phi'=0$ is a natural choice to have a nonsingular solution. In the case with gravity, we need further conditions on $\rho$. One may choose the initial condition type of boundary conditions $\rho=0$ and $\rho'={\rm const}$  at $\eta=0$.

These boundary conditions will give small bubbles within large
backgrounds regardless of the background geometry \cite{bl03}. On the other hand, the theory has a wide variety of solution types if the effect of gravity becomes more significant. Large bubbles have large values of the cosmological constant, while small bubbles have small values of the cosmological constant \cite{bl03}.

In the case of Euclidean de Sitter background with compact geometry, the following choice of boundary conditions is also possible: $\Phi'=0$ and $\rho=0$ at $\eta=0$ and $\eta_{max}=0$ \cite{hw02}. This boundary condition has $Z_2$ symmetry under the exchange of two points corresponding to $\eta=0$ and $\eta_{max}=0$. In general, for a potential with degenerate vacua, we can choose the boundary condition with $Z_2$ symmetry as follows:
\begin{equation}
\frac{d\Phi}{d\eta}\Big|_{\eta=0}=0,
\,\,\,\, \frac{d\Phi}{d\eta}\Big|_{\eta=max}=0, \,\,\,\,
\rho|_{\eta=0}=0 , \,\,\,\, {\rm and}\,\,
\rho|_{\eta=\eta_{max}}=\tilde\rho(\eta_{max})=0, \label{ourbc-2}
\end{equation}
where $\eta_{max}$ will have a finite value in
the cases we will consider. We adopt this boundary condition for our instanton solutions. $\Phi(\eta_{max})$ is exponentially
approaching to but not reaching $\Phi^v_f$. The above boundary conditions have been used in dS background space \cite{hw02} as already mentioned. We will also choose $\tilde\rho(\eta_{max})=0$ for both flat and AdS space to impose $Z_2$ symmetry. This will allow new solutions with $Z_2$ symmetry even both flat and AdS space. Because of $Z_2$ symmetry about the wall, the inside geometry will be identical to the outside one. As a result, the whole geometry is finite. This is very different from the known solutions with infinite geometry \cite{col002, bnu02, par02, hiscock}.

\begin{figure}
\begin{center}
\includegraphics[width=2.0in]{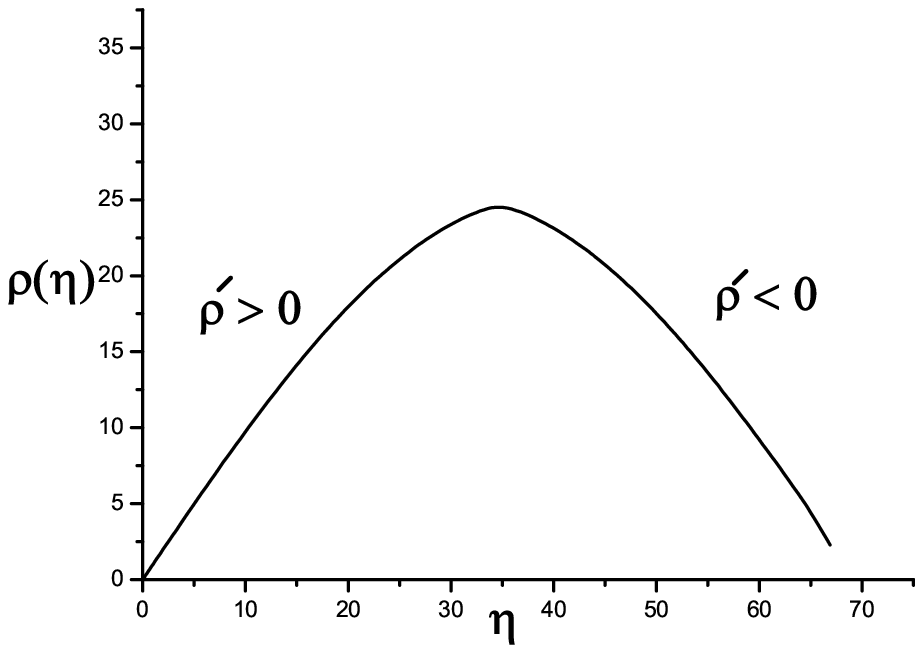}
\includegraphics[width=2.0in]{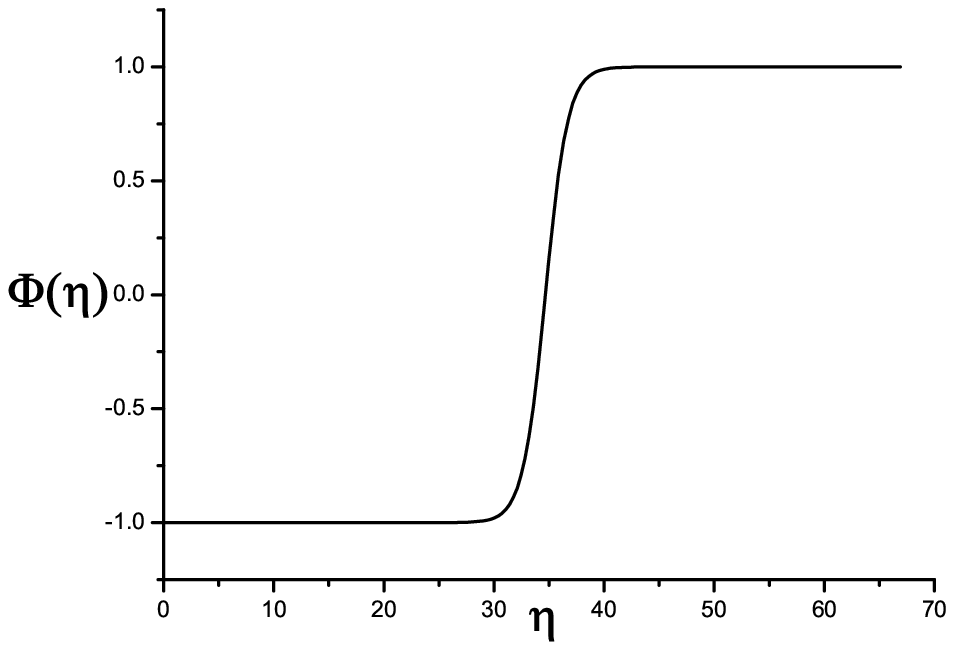}
\includegraphics[width=2.0in]{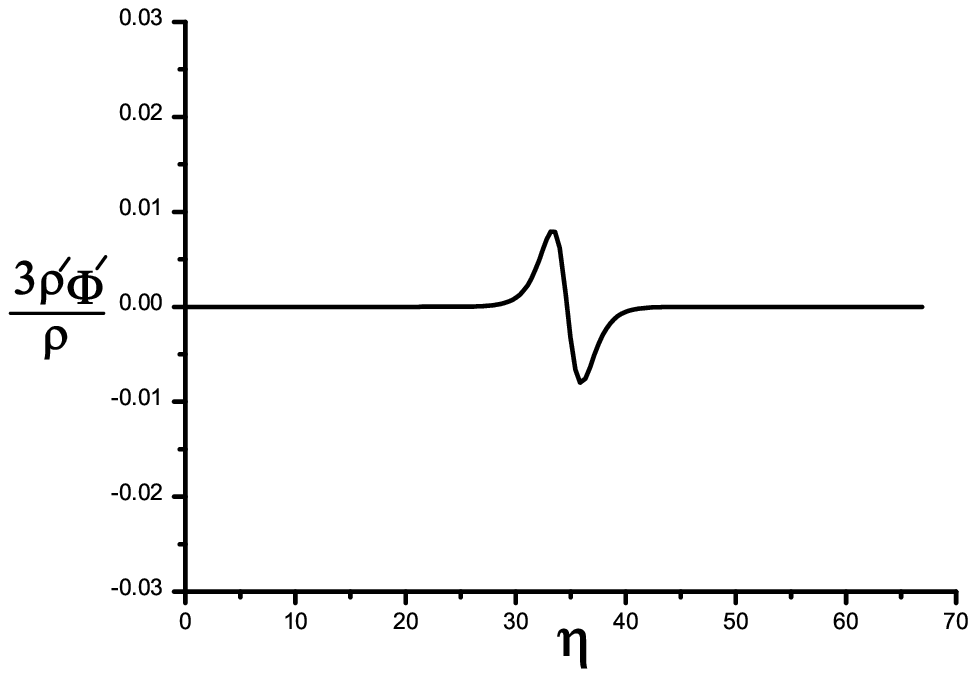}\\
(A) Instanton solution in de Sitter space\\
\includegraphics[width=2.0in]{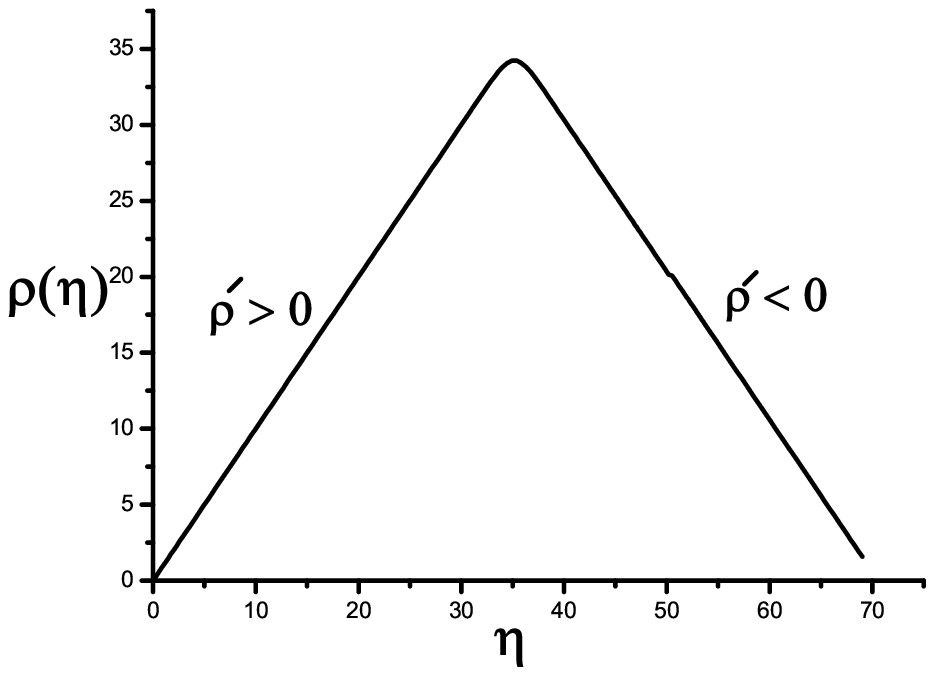}
\includegraphics[width=2.0in]{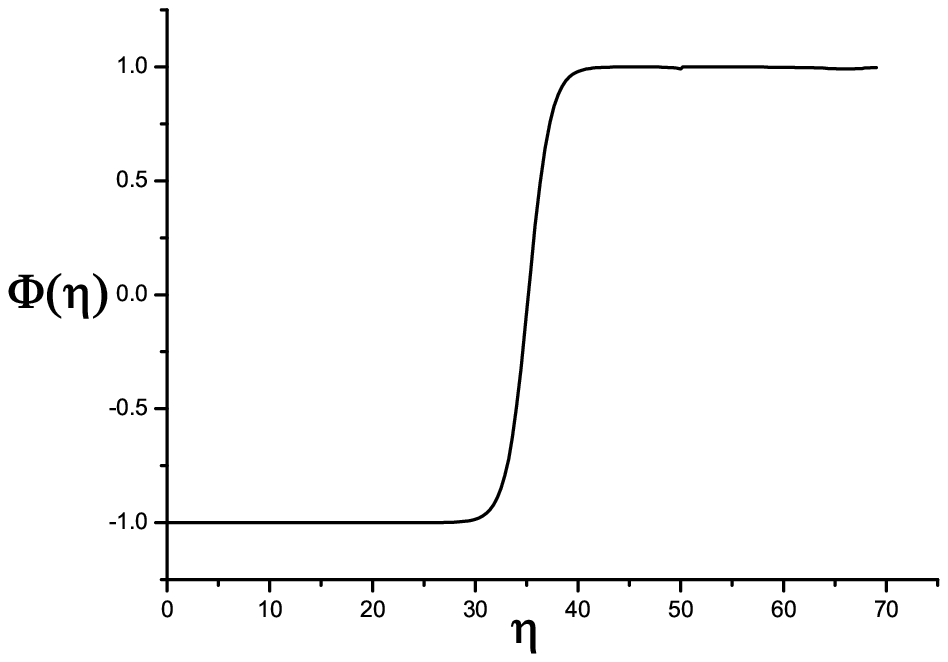}
\includegraphics[width=2.0in]{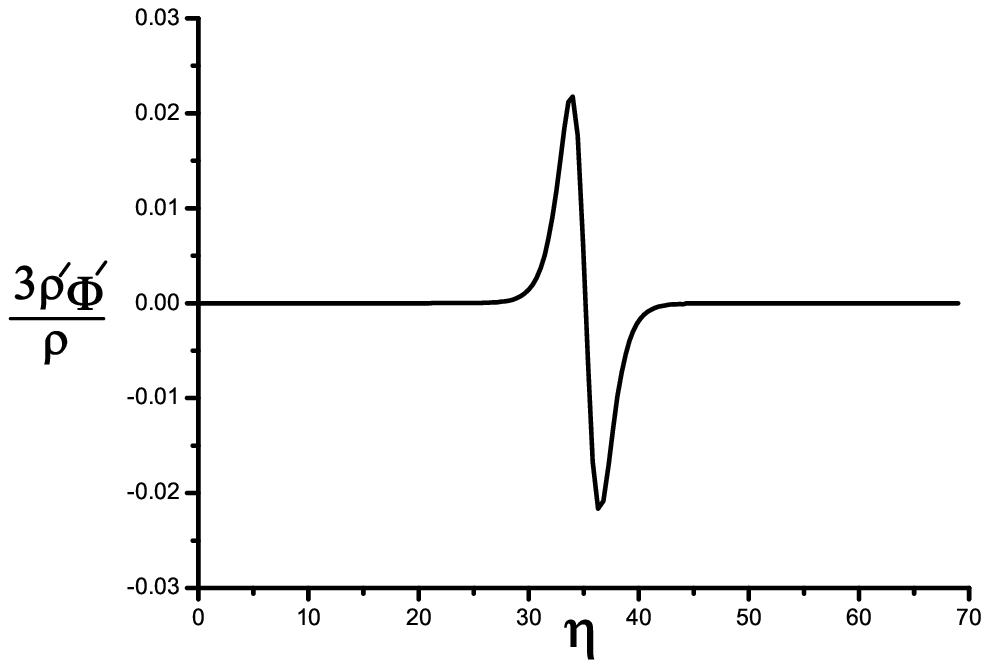}\\
(B) Instanton solution in flat space\\
\includegraphics[width=2.0in]{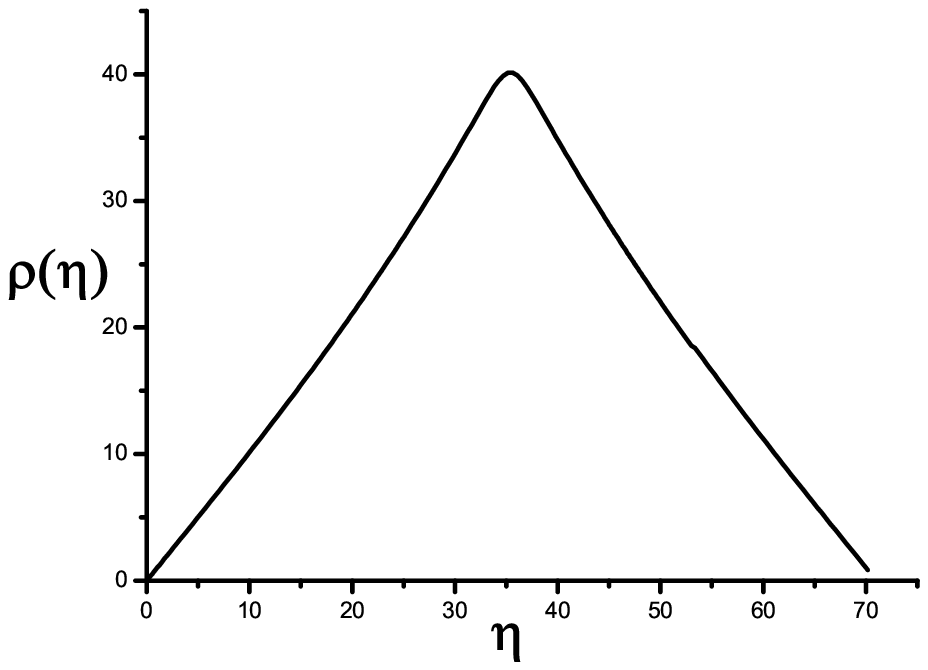}
\includegraphics[width=2.0in]{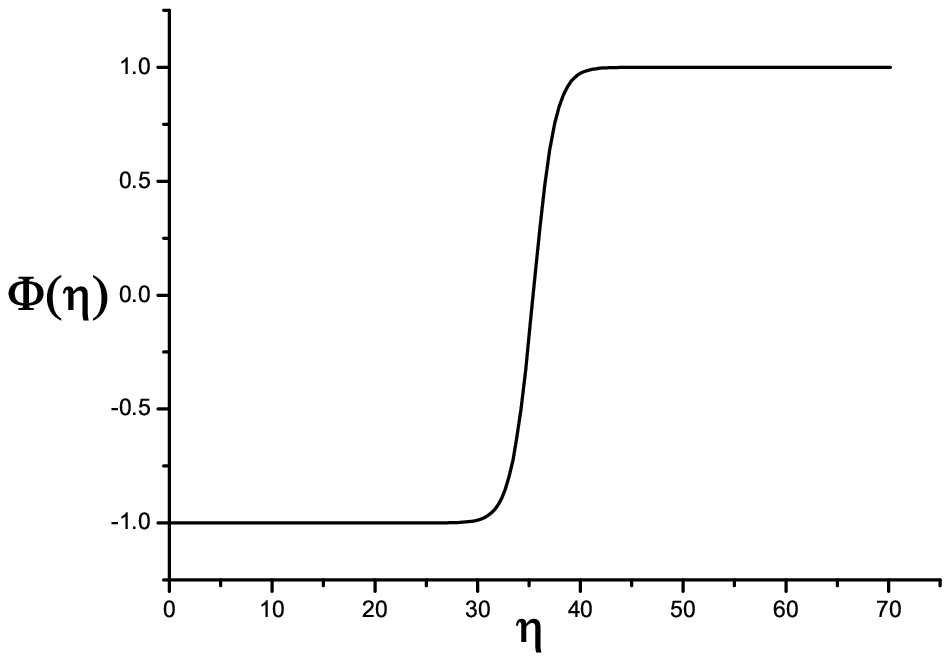}
\includegraphics[width=2.0in]{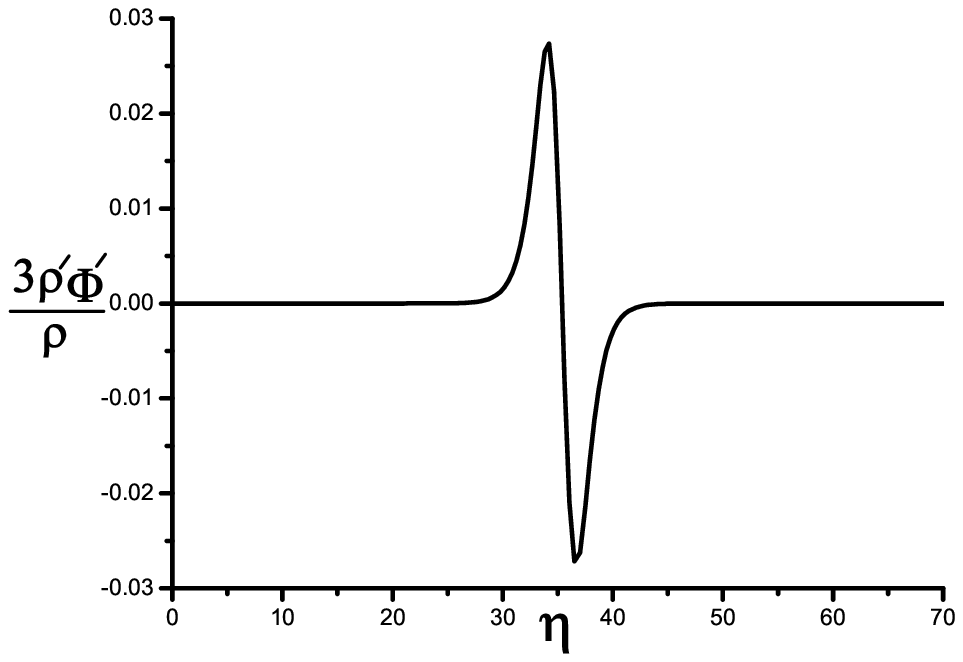}\\
(C) Instanton solution in anti-de Sitter space\\
\end{center}
\caption{\footnotesize{The first and second columns show the
numerical solutions on $\rho$ and $\Phi$, respectively. The third
column shows the evolution of the second term in the left one of
Eq.\ (\ref{ephi}). Row (A) illustrates the transition
between the degenerate vacua in dS space, $\hat{U}_o=0.01$ and
$\hat{\kappa}\simeq 0.0474$. Row (B) illustrates the
transition between the degenerate vacua in flat space,  $\hat{U}_o=0$ and $\hat{\kappa}\simeq 0.1805$. Row (C) illustrates the
transition between the degenerate vacua in AdS space, $\hat{U}_o=-0.01$ and
$\hat{\kappa}\simeq 0.2392$. }} \label{fig:fig01}
\end{figure}

In what follows, we explore in more detail the possibility for the existence of the instanton solution. To understand solutions qualitatively, we rearrange the terms in Eq.\ (\ref{ephi}) after multiplying by $\frac{d\Phi}{d\eta}$
\begin{equation}
\int^{\eta_{max}}_{0} d \left[\frac{1}{2}\Phi'^{2}-U \right]
 = - \int^{\eta_{max}}_{0} d\eta \frac{3\rho'}{\rho}\Phi'^{2}.  \label{econser}
\end{equation}
The quantity in the square brackets here can be interpreted as the
total energy of the particle with the potential energy $-U$. The
term on the right-hand side can be considered as the dissipation
rate of the total energy as long as $\rho' > 0 $. However, the role of the term can be changed from damping to acceleration if $\rho'$
changes sign during the transition. For the tunneling between
the degenerate vacua the region for $\rho' > 0$ and $\rho' < 0$
will be equally divided due to the $Z_2$ symmetry, hence the term
on the right-hand side will be vanished. Thus, the total energy at both ends, $\eta=0$ and $\eta=\eta_{max}$, is conserved as we can see from Eq.\ (\ref{econser}) and the third column in Fig.\ \ref{fig:fig01}. The condition for the change of
sign depends on the maximum value of the potential. To allow
the change during the transition, the local maximum value of the
potential $U(0)$ must be positive. This can be seen from Eq.\ (\ref{erho}). That is to say,
\begin{equation}
U(0) = \frac{3}{\kappa\rho^2} + \frac{1}{2}\Phi'|^2_{(\Phi=0)} > 0. \nonumber
\end{equation}
In other words, the sign of the second term of the first equation in Eq.\ (\ref{ephi}) can be changed during the transition for $-\mu^4/8\lambda < U_o$.

The numerical solutions for the equations are illustrated in Fig.\
\ref{fig:fig01}. In this work, we employ the following dimensionless variables \cite{wl01}:
\begin{equation}
\hat{U}(\hat{\Phi})= \frac{\lambda U(\Phi)}{\mu^4},\;\; \hat{\Phi^2}=\frac{\lambda \Phi^2}{\mu^2},\;\;\hat{\eta}=\mu\eta,\;\;\hat{\rho}=\mu\rho,\;\; \hat{\kappa}=\frac{\mu^2\kappa}{\lambda}.
\end{equation}
These variables give
\begin{equation}
\hat{U}(\hat{\Phi})=\frac{1}{8}\left(\hat{\Phi}^2 -1\right)^2+ \hat{U}_o,
\end{equation}
and the Euclidean field equations for $\Phi$ and $\rho$ become
\begin{equation}
\hat{\Phi}'' + \frac{3\hat{\rho}'}{\hat{\rho}}\hat{\Phi}'=\frac{d\hat{U}}{d\hat{\Phi}} \,\,\,\, {\rm
and} \,\,\,\, \hat{\rho}'' = - \frac{\hat{\kappa}}{3}\hat{\rho} (\hat{\Phi}'^2 +\hat{U}),
\end{equation}
respectively and the Hamiltonian constraint is given by
\begin{equation}
\hat{\rho}'^2 - 1 -
\frac{\hat{\kappa}\hat{\rho}^2}{3}\left(\frac{1}{2}\hat{\Phi}'^{2}-\hat{U}\right) = 0 .
\end{equation}
We use relaxation methods to solve Eqs.\
(\ref{ephi}) and (\ref{erho}) with boundary conditions
(\ref{ourbc-2}). For dS space, this solution was already studied
in Refs.\ \cite{hw02, bl03}. In Fig.\ \ref{fig:fig01}, the first row (A) illustrates the
transition between the degenerate vacua in dS space, $U_o > 0$.
The second row (B) illustrates the transition between the
degenerate vacua in flat space. The third row (C) illustrates the
transition between the degenerate vacua in AdS space. The first
column corresponds to the solution of $\rho$. The second column
corresponds to the solution of $\Phi$. The third column
illustrates the evolution of the second term on the left side of
Eq.\ (\ref{ephi}). In the first column in Fig.\ \ref{fig:fig01}, we can see that the role of the second term on the left side of Eq.\ (\ref{ephi}) changes from damping $\rho' > 0$ to acceleration $\rho' < 0$ during the transition. For the size of $S^3$, $\rho$ starts from zero at $\eta=0$, reaches the maximum size at $\eta=\eta_{max}/2$, and becomes zero again at $\eta=\eta_{max}$. Obviously, the geometry is finite and $Z_2$ symmetry. Note that $\tilde\rho(\eta_{max})$ goes to zero unlike the bounce solutions where $\tilde\rho(\eta_{max})$
goes to infinity for both flat and AdS space.

We must be careful in evaluating the action difference between the action of the solution and the background action itself. In previous works \cite{col002, bnu02, par02}, the outside contribution to the action of the solution is equal to that from outside of the background. In our solution, the outside geometry of the solution is quite different from that of the background.
We can not employ Eq.\ (3.8) in Ref.\ \cite{bnu02} directly because the role of the second term $\rho'$ is changed from the initial role of damping to the finial role of acceleration. When we do integration by parts in an analytic computation we cannot simply drop out the surface term due to our boundary conditions Eq.\ (\ref{ourbc-2}).

In the thin-wall approximation scheme, the Euclidean action can be
divided into three parts: $\Delta S= \Delta S_{in} + \Delta S_{wall} + \Delta S_{out}$, where
$\Delta S=S_E(solution)-S_E(background)$. For flat and AdS cases the
inside part $\Delta S_{in}$ gives no contribution because of unchanged inside geometry in the tunneling solution and background. Here, we will send the contribution of the inside part due to the diminished inside bulk shape to the contribution of the outside part for simplicity. Thus, we only need to consider the contribution of the wall and the outside part. The contribution of the wall is given by $\Delta S_{wall}=2\pi^2 \bar{\rho}^3 S_o$, where the surface tension of the wall $S_o(=2\mu^3/3\lambda)$ is a constant \cite{bnu02}. We note that it corresponds to the energy of a kink in one-dimensional space in the form of the potential without $U_o$ in Eq.\ (\ref{poten}). The contribution from the outside part is given by $\Delta S_{out} = S_E(solution)_{out}-S_E(background)_{out}$.

In the thin-wall approximation scheme, the relation between $d\rho$ and $d\eta$ can be seen by the relation
\begin{equation}
d\rho =\pm d\eta \left[1-\frac{\kappa\rho^2U}{3}\right]^{1/2},
\label{lee001-inw}
\end{equation}
where $+$ is for $0 \le \eta < \eta_{max}/2$, $0$ is for
$\eta=\eta_{max}/2$, and $-$ is for $\eta_{max}/2 < \eta \le
\eta_{max}$ \cite{bl03}.

Outside geometry after tunneling will be finite as that of dS
space. Therefore we will consider the initial background space
having the size $\tilde\rho^i_{max}(\eta^i_{max})$. $\eta^i_{max}$
denotes $\eta_{max}$ in the initial space. The finite size in the
initial space will go to infinity as
$\tilde\rho^i_{max}(\eta^i_{max})$ goes to infinity except for dS
space. We also study the effect as the size increases. For both
flat and AdS space, we have to carry out integration by parts
carefully. The contribution from integration by parts Eq.\ (3.8)
in Ref.\ \cite{bnu02} gives as follows
\begin{equation}
S_{E}(background)_{out} = 4\pi^2 \int^{\eta^i_{max}}_{\bar{\eta}} \left( \rho^3
U- \frac{3\rho}{\kappa} \right) d\eta - \Delta S_{ibp},
\end{equation}
where the second term on the right-hand side of the above equation
is from the effect of the surface term at $\tilde\rho^{i}_{max}$
and $\Delta S_{ibp} = - (6\pi^2/\kappa)\tilde\rho^{i2}_{max}\left( 1-\kappa\tilde\rho^{i2}_{max} U_o/3 \right)^{1/2}$. For dS space,
$\rho^{i}_{max}(\eta_{max})=0$ and $\Delta S_{ibp} = 0$.

The analytic computation in dS space was obtained \cite{bl03} as follows:
\begin{equation}
\Delta S_{bulk}= \Delta S_{in} +\Delta S_{out} =  \frac{12\pi^2}{\kappa^2 U_o} \left
(1-\frac{\kappa}{3}U_o\bar{\rho}^2\right)^{3/2} +
\frac{12\pi^2}{\kappa^2 U_o} \left
(1-\frac{\kappa}{3}U_o\bar{\rho}^2\right)^{3/2}. \label{a-bulk}
\end{equation}
The first term is from the contribution of the inside bulk part due to the diminished inside bulk shape and the second term is from the
outside bulk part due to the same reason. Then, the coefficient
$\Delta S$ and the critical radius of the wall are evaluated to be
\begin{equation}
\bar{\rho} = \frac{2}{\kappa\sqrt{\frac{S^2_o}{4} +
\frac{4}{3\kappa}U_o}} \;\;\; {\rm and}\;\;\; \Delta S = \frac{12\pi^2
S_o}{\kappa^2 U_o\sqrt{\frac{S^2_o}{4} + \frac{4}{3\kappa}U_o}},  \nonumber
\end{equation}
where Eq.\ (\ref{lee001-inw}) is used.

The final form from the contribution of the outside part in both flat and AdS space is evaluated to be
\begin{equation}
\Delta S_{out} = \frac{12\pi^2}{\kappa^2U_o}\left[2\left( 1-
\frac{\kappa}{3}\bar\rho^2U_o\right)^{3/2} - 1 - \left( 1-
\frac{\kappa}{3}\tilde\rho^{i2}_{max} U_o \right)^{3/2}\right] +
\Delta S_{ibp}. \label{fads-out}
\end{equation}
The minus sign in front of the third term in the square
parentheses is owing to the integration of the outside part in flat
and AdS space. For dS space, the minus sign is changed into the
plus. This is because Eq.\ (\ref{lee001-inw}) is used for both the
tunneling solution and the background in evaluating $\Delta S_{out}$, while Eq.\ (\ref{lee001-inw}) is used only for the tunneling solution in both flat and AdS cases. Equation\ (\ref{fads-out}) with the plus
sign and $\tilde\rho^{i}_{max}(\eta^i_{max})=0$ will reproduce the result for dS in Eq.\ (\ref{a-bulk}).

The position of the wall obtained by extremizing $\Delta S$ is given by
\begin{equation}
\bar{\rho} = \frac{2}{\kappa\sqrt{\frac{S^2_o}{4} +
\frac{4}{3\kappa}U_o}}. \label{crw02}
\end{equation}
This form has the same one as the tunneling in dS space
\cite{bl03} except that $U_o < 0$. The transition rate is evaluated
to be
\begin{equation}
\Delta S = \frac{12\pi^2}{\kappa^2U_o}\left[
\frac{S_o}{\sqrt{\frac{S^2_o}{4} + \frac{4}{3\kappa}U_o}} - 1 -
\left( 1- \frac{\kappa}{3}\tilde\rho^{i2}_{max} U_o
\right)^{3/2}\right] + \Delta S_{ibp}.
\end{equation}

We first consider the case for $1 < \frac{\kappa}{3}\tilde\rho^{i2}_{max}|U_o|$. As the initial value of $\tilde\rho^{i}_{max}$ goes to infinite, the exponent becomes
\begin{equation}
\Delta S \sim -2\pi^2 \sqrt{\frac{|U_o|}{3\kappa}} \tilde\rho^{i3}_{max}. \nonumber
\end{equation}
As the initial space goes to infinite, the transition rate has an exponentially growing value.

We now consider the case for $1 > \frac{\kappa}{3}\tilde\rho^{i2}_{max}|U_o|$. Then the final form of $\Delta S$ is evaluated to be
\begin{equation}
\Delta S \sim - \frac{64\pi^2}{\kappa^3 S^2_o}-\pi^2\tilde\rho^{i2}_{max}|U_o|, \nonumber
\end{equation}
where $|U_o|$ is equal to zero for the case of flat space. The exponent $\Delta S$ has the negative minimum value at $\rho=\bar{\rho}$. The action of the solution is less than the action of the background for the cases in both flat and AdS space. For the case in dS space, the action difference has the positive value, $\Delta S > 0$.

From the above results, the instanton solution in dS space can be interpreted as the ordinary formation process of a kink or domain wall. On the other hand, the solution in both flat and AdS space can be interpreted as the mixed state of the two degenerate vacuum states. We expect that this phenomenon may be interpreted as the analog of the energy difference, $\Delta E < 0$, between the energy of the mixed state and the harmonic oscillator in the left or right potential.

\section{The dynamics of the solution in Lorentzian spacetime  \label{sec3}}

\begin{figure}
\begin{center}
\includegraphics[width=3.in]{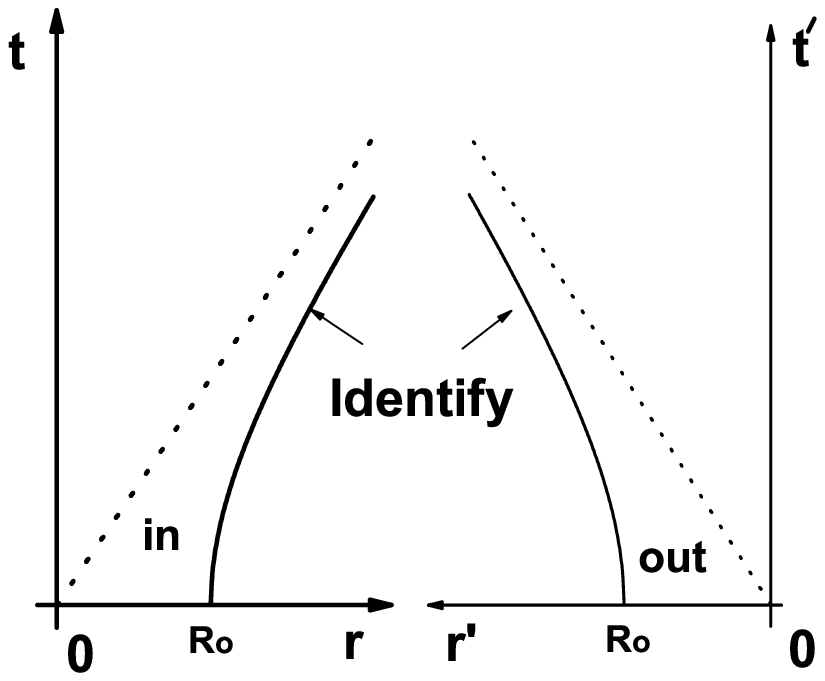}
\includegraphics[width=3.in]{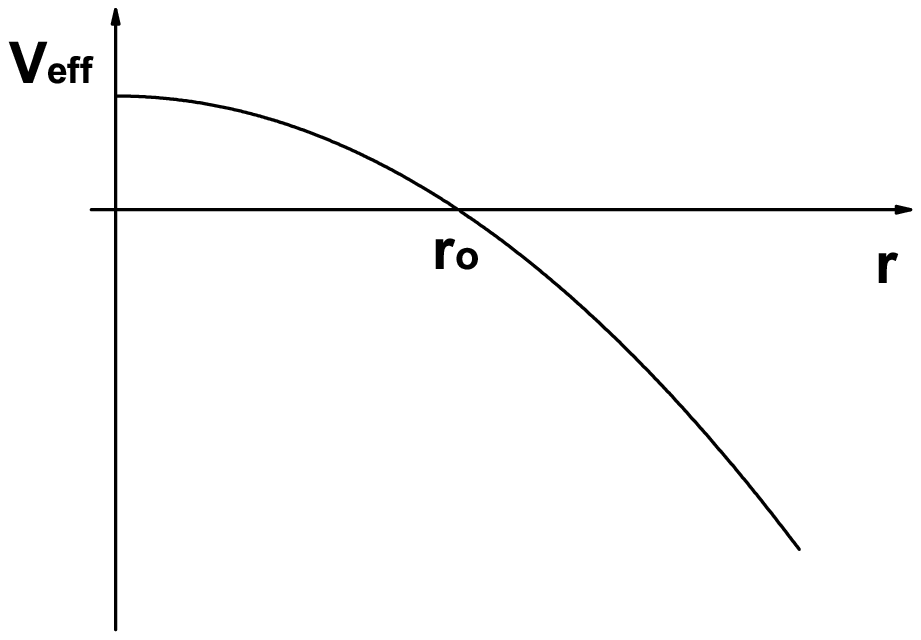}
\end{center}
\caption{\footnotesize{The figures illustrate the time evolution
after the nucleation. The left figure represents the evolution of the wall under the inside and outside observers's point of view.
The right figure represents the effective potential in the junction equation.}}
\label{fig:fig02}
\end{figure}

In this section, we study the growth of the wall after its
materialization. To obtain the dynamics of the solutions in
Lorentzian spacetime, we now apply the analytic continuation as in
Ref.\ \cite{chw02}
\begin{equation}
\chi = i \xi + \pi/2,
\end{equation}
where $\pi/2$ is added for a proper Lorentzian signature
$(-,+,+,+)$. The only difference is that one has to
continue the scalar field as well as $O(4)$-invariant
Euclidean space into an $O(3,1)$-invariant Lorentzian spacetime.
This analytic continuation gives the
spherical Rindler-type metric \cite{gerlach} as follows:
\begin{eqnarray}
ds^2 &=& d\eta^2 + \eta^2 [-d\xi^2 +
\cosh^2\xi(d\theta^2+\sin^2\theta
d\phi^2)] \;\;\; {\rm for \;\; flat},  \label{ct001} \\
ds^2 &=& d\eta^2 +
\frac{3}{\Lambda}\sinh^2\sqrt{\frac{\Lambda}{3}}\eta [-d\xi^2 +
\cosh^2\xi(d\theta^2+\sin^2\theta d\phi^2)] \;\;\; {\rm for \;\;
AdS},  \label{ct06} \\
ds^2 &=& d\eta^2 +
\frac{3}{\Lambda}\sin^2\sqrt{\frac{\Lambda}{3}}\eta [-d\xi^2 +
\cosh^2\xi(d\theta^2+\sin^2\theta d\phi^2)] \;\;\; {\rm for \;\;
dS}, \label{ct60}
\end{eqnarray}
where $\Lambda=\kappa |U_o|$ and $\xi$ becomes the timelike
coordinate. In Minkowski spacetime [see Eq.\ (\ref{ct001})], the inside geometry has the relation $\eta^2=-t^2+r^2$ for the range of $0\leq \eta < \eta_{max}/2$. On the other hand, the outside geometry has the relation $\tilde{\eta}^2=-t^2+r^2$ for the range of $\eta_{max}/2 < \eta \leq \eta_{max}$, where we defined $\tilde{\eta}=\eta_{max}-\eta$. The geometry can be constructed by joining two spacetimes along the wall, $\bar{\eta} =\eta_{max}/2$. As a result, the spacetime has $Z_2$ symmetry. At the initial Lorentzian time ($t=0$), $\bar{\eta}$ is equal to $R_o$ as shown in the left figure in Fig.\ \ref{fig:fig02}. The method obtaining the left
figure in Fig.\ \ref{fig:fig02} was studied in Ref.\ \cite{col002}.

The geometry and properties of our solutions as compared with the domain wall \cite{dvil, dip} were studied in Ref.\ \cite{bl03}. Domain walls can form in any model having a spontaneously broken discrete symmetry. An inertial observer sees the domain wall accelerating away with a specific acceleration. The domain wall has repulsive gravitational fields \cite{dvil, dip}. On the other hand, our solutions are different from the formation process of a domain wall. The spacetime in the present work depends only on the cosmological constant and also has the spherical Rindler-type metric.

Now we will employ the Israel junction condition. In the thin-wall limit, all the energy momentum is localized on the wall and therefore the thin wall has a surface stress-energy tensor. Thus, it becomes a kind of surface layer \cite{isr01}. In this framework, only the vacuum Einstein equations should be solved on either side of the wall. We use another coordinate transformation \cite{bl03, chw02} for $O(3)$ invariant configurations:
\begin{eqnarray}
R&=& \eta \cosh\xi, \;\;\; T = \eta\sinh\xi, \;\;\; {\rm for \;\; flat} \nonumber \\
R&=& \sqrt{\frac{3}{\Lambda}}\sinh\sqrt{\frac{\Lambda}{3}}\eta\cosh\xi,\;\;\; T = \sqrt{\frac{3}{\Lambda}}\tan^{-1} \left(\sinh\xi \tanh\sqrt{\frac{\Lambda}{3}}\eta \right), \;\;\; {\rm for \;\; AdS}  \label{ct04} \\
R&=& \sqrt{\frac{3}{\Lambda}} \sin\sqrt{\frac{\Lambda}{3}}\eta\cosh\xi,
\;\;\; T = \frac{1}{2}\sqrt{\frac{3}{\Lambda}} \ln \frac{\cos\sqrt{\Lambda/3}\eta+\sin\sqrt{\Lambda/3}\eta\sinh\xi} {\cos\sqrt{\Lambda/3}\eta -\sin\sqrt{\Lambda/3}\eta\sinh\xi}, \;\;\; {\rm for \;\;
dS}.   \nonumber
\end{eqnarray}
Because of the exact $Z_2$ symmetry of the whole spacetime, both inside and outside have the same behavior.
The metric in Eqs.\ (\ref{ct001}, \ref{ct06}, \ref{ct60}) becomes
\begin{equation}
ds^2 = - \left( 1 \pm \frac{\Lambda}{3} R^2 \right) dT^2 +
\frac{dR^2}{\left( 1 \pm \frac{\Lambda}{3} R^2 \right)} + R^2
d\Omega^2_2 . \label{o3metric}
\end{equation}
We introduce the Gaussian normal coordinate system near the wall
\begin{equation}
dS^2 = - d\tau^2 + d\eta^2 + {\bar r}^2(\tau, \eta) d\Omega^2,
\end{equation}
where $g_{\tau\tau}=-1$ and ${\bar r}(\tau, \bar{\eta})=r(\tau)$.
It must agree with the coordinate $R$ of the left and right regions. The angle variables can be taken to be
invariant in all regions. The induced metric on the wall is given by
\begin{equation}
dS^2_{\Sigma} = - d\tau^2 + r^2(\tau) d\Omega^2,
\end{equation}
where $\tau$ is the proper time measured by an observer at rest
with respect to the wall and $r(\tau)$ is the proper radius of
$\Sigma$. The metrics in both sides of the wall give the same induced metric on the wall. The following condition should be satisfied: $1=(1\pm\Lambda R^2/3)\dot{T}^2-(1\pm\Lambda R^2/3)^{-1} \dot{R}^2$, where the dot is a derivation with respect to $\tau$. Because of the spherical symmetry, the extrinsic curvature has only two
components, $K^{\theta}_{\theta} \equiv K^{\phi}_{\phi}$ and
$K^{\tau}_{\tau}$. The junction equation is related to
$K^{\theta}_{\theta}$ and the covariant acceleration in the normal
direction is related to $K^{\tau}_{\tau}$. The junction condition then becomes
\begin{equation}
\sqrt{1 \pm \frac{\Lambda}{3}r^2 + \dot{r}^2} + \sqrt{1 \pm
\frac{\Lambda}{3}r^2 + \dot{r}^2} = \frac{\kappa}{2}\sigma r,
\end{equation}
where $\sigma$ is the surface-energy density and equivalent to the surface tension of the wall $S_o$. After squaring, we obtain the metric junction equation. The effective potential has the form
\begin{equation}
V_{eff}=\frac{1}{2} \left[- \frac{\kappa^2}{16}\sigma^2 \pm
\frac{\Lambda}{3}\right]r^2 + \frac{1}{2}.
\end{equation}
The condition for the existence of the expanding solution is $\Lambda < 3\kappa^2\sigma^2/16$ for AdS spacetime, while the expanding solution is always possible for dS spacetime.

The left figure in Fig.\ \ref{fig:fig02} represents the time evolution of the inside and outside spacetime from the inside and outside observers's point of view, respectively. The location of walls is denoted by $R_o$. Because the configuration of the whole spacetime has $Z_2$ symmetry the walls may be identified. Then both observers in the left and right sides of the wall feel that they are inside of the wall. The right figure represents the effective potential describing the dynamics of the wall under the metric junction equation. We can read off directly the properties of the trajectory of the wall from the shape of the effective potential. The location of $r_o$ where $V_{eff}=0$ in the right figure in Fig.\ \ref{fig:fig02} is given by
\begin{equation}
r_{o}=\frac{2}{\kappa\sqrt{\frac{\sigma^2}{4}\mp\frac{4\Lambda}{3\kappa^2}}}. \end{equation}
This $r_o$ is the same as the position of the wall in Eq.\ (\ref{crw02}). After the tunneling transition, the wall can expand
without eating up bulk (inside and outside or left and right) spacetime.

\section{Summary and Discussions}

In this paper, we have studied instanton solutions with $O(4)$ symmetry between the degenerate vacua in curved space. The boundary conditions we imposed have $Z_2$ symmetry, which are different from the usual ones. The solution also has the exact $Z_2$ symmetry and gives rise to the geometry of a finite size. We obtained the numerical solutions as well as performed the analytic computations using the thin-wall approximation. These nontrivial solutions are possible only if gravity is taken into account. The solutions which represent the tunneling in the opposite direction, anti-instanton, are easily obtained by $\eta \rightarrow \eta_{max} -\eta$. The leading semiclassical exponent $\Delta S$ is given by the instanton solution. To evaluate $\Delta S$ we considered the initial background space with the finite size $\tilde\rho^i_{max}$ cutoff. The finite size in the initial space will go to infinity as $\tilde\rho^i_{max}$ goes to infinity except for dS space. We studied the effect as the size increases. In the thin-wall approximation, the solution describing tunneling in flat space has the negative minimum value of the exponent $\Delta S$. This will be the dominant contribution to the Euclidean path integral. For AdS space, the transition rate has a negative exponentially growing value due to the diverging background subtraction. For dS space, the
coefficient $\Delta S$ has a positive value. The difference among the dS, flat, and AdS space is caused by the initial size of the
background Euclidean space or the diverging background subtraction
depending on the cosmological constant. The instanton solution in dS space can be interpreted as the ordinary formation process of a kink or domain wall. On the other hand, the solution in both flat and AdS space can be interpreted as the mixed state of the two degenerate vacuum states. We expect that this phenomenon may be interpreted as the analog of the energy difference, $\Delta E < 0$, between the energy of the mixed state and the harmonic oscillator in the left or right potential. It is not clear how to interpret the physical meaning of the negative value of $\Delta S$ in the present work. However, the exponent $\Delta S$ can have a finite value if the tunneling describing our instanton solutions occurs in the initial background with finite size.

In these solutions mediating tunneling between the degenerate
vacua, the observers's point of view is different from the ordinary
formation process of the vacuum bubble or domain wall. Our instanton solutions have specific dynamics in Lorentzian spacetime. In this formation, there is only one point of view as inside. For example, every observer sees themselves surrounded by the wall after the tunneling in dS space. In other words, the wall is located at $R_o(=\rho_{max}(\eta_{max}/2))$. The inside and the outside observers are located at $r=0$ and $r'=0$, respectively. After the analytic continuation from Euclidean to Lorentzian the observer is inside of the wall and the other is also inside of the wall over the dS horizon. Thus, every observer remains inside of the wall. The situation will be similar in both flat and AdS space although they do not have a real horizon. The wall formed when the particle rolls down and up the inverted potential in the mechanical analogy. The middle of the wall has the energy density corresponding to the top of the potential, which is equivalent to the vacuum energy of dS space. Thus, the wall somehow plays the role of the dS horizon, but not the real horizon. The other observer exists over the wall, which is located at $r'=0$. It seems that the solutions represent 2-brane formation with $Z_2$ symmetry. Thus, the topology of the initial spacetime could be changed under the influence of this solution.

In order to get the analytic computation on the action
one may consider carefully the contribution from the
Gibbons-Hawking term. This is because the geometry of the outside wall of our solutions is different from that of the background. This is not needed for the usual bubble solutions where the outside geometry does not change. We expect that the point $\tilde{\eta}=0$ after the tunneling process is smooth due to $Z_2$ symmetry.

We propose the solutions with exact $Z_2$ symmetry can represent
the nucleation of the braneworld-like object if the mechanism is
applied in higher-dimensional theory. The braneworld having the
finite size with the exact $Z_2$ symmetry can be nucleated not
only in dS but also in flat and AdS bulk spacetime, and then expand,
seen from an observer's point of view on the wall, without eating up
bulk (inside and outside) spacetime.

It will be interesting to see if this type of solution can be extended to the theories with gauge fields or in various dimensions. These solutions are interpreted as an instanton solution rather than a bounce solution in not only dS but also flat and AdS space. The
study on the instanton solution in curved space may be important
for the understanding of the properties of the vacuum structure in
these theories as well as to see if the processes may happen in the early universe. Can we obtain the situation describing each observer as an outside one? It may be related to the wormhole. The study
including a wormhole solution will also be interesting.

\section*{Acknowledgements}

We would like to thank E. J. Weinberg, M. B. Paranjape, Sang-Jin
Sin, and Hongsu Kim for helpful discussions and comments. We would
like to thank Kei-ichi Maeda for helpful discussions at the 18th
Workshop on General Relativity and Gravitation in Japan (JGRG18).
This work was supported by the Korea Science and Engineering
Foundation (KOSEF) grant funded by the Korea government(MEST)
through the Center for Quantum Spacetime(CQUeST) of Sogang
University with grant number R11 - 2005 - 021. W.L. was supported
by the Korea Research Foundation Grant funded by the Korean
Government(MOEHRD) (KRF-2007-355-C00014).


\newpage

\begin{thebibliography}{99}
\bibitem{hw02} J. C. Hackworth and E. J. Weinberg, Phys. Rev. D {\bf 71}, 044014 (2005).
\bibitem{bl03} B.-H. Lee and W. Lee, Classical Quantum Gravity {\bf 26}, 225002 (2009).
\bibitem{voloshin} M. B. Voloshin, I. Yu. Kobzarev, and L. B.
Okun, Yad. Fiz. {\bf 20}, 1229 (1974)[Sov. J. Nucl. Phys. {\bf
20}, 644 (1975)].
\bibitem{col002} S. Coleman, Phys. Rev. D {\bf 15}, 2929 (1977); {\it ibid.} D {\bf 16}, 1248(E) (1977).
\bibitem{bnu02} S. Coleman and F. De Luccia, Phys. Rev. D {\bf 21}, 3305 (1980).
\bibitem{par02} S. Parke, Phys. Lett. {\bf 121B}, 313 (1983).
\bibitem{bpst} A. Belavin, A Polyakov, A Schwartz, and Y. Tyupkin, Phys. Lett. {\bf 59B}, 85 (1975); G. 'tHooft, Phys. Rev. D {\bf 14}, 3432 (1976); Phys. Rev. Lett. {\bf 37}, 8 (1976).
\bibitem{ggp} G. Gibbons, M. Green, and M. Perry, Phys. Lett. B {\bf 370}, 37 (1996); A. V. Belitsky, S. Vandoren, and P. van Nieuwenhuizen, Classical Quantum Gravity {\bf 17}, 3521 (2000).
\bibitem{vni} S. Vandoren and P. van Nieuwenhuizen, arXiv:0802.1862.
\bibitem{ghh} S. W. Hawking, Phys. Lett. {\bf 60A}, 81 (1977); G. W. Gibbons and S. W. Hawking, Phys. Lett. {\bf 78B}, 430 (1978); G. 'tHooft, Nucl. Phys. {\bf B315}, 517 (1989).
\bibitem{page} D. N. Page, arXiv:0912.4922.
\bibitem{hb0} E. Baum, Phys. Lett. {\bf 133B}, 185 (1983); S. W. Hawking, Phys. Lett. {\bf 134B}, 403 (1984); A. D. Linde, JETP {\bf 60}, 211 (1984); A. Vilenkin, Phys. Rev. D {\bf 30}, 509 (1984).
\bibitem{vhl} S. W. Hawking and N. Turok, Phys. Lett. B {\bf 425}, 25 (1998); {\it ibid.} B {\bf 432}, 271 (1998); A. Vilenkin, Phys. Rev. D {\bf 57}, R7069 (1998); A. Linde, Phys. Rev. D {\bf 58}, 083514 (1998).
\bibitem{ghp000} G. W. Gibbons, S. W. Hawking, and M. J. Perry, Nucl. Phys. {\bf B138}, 141 (1978).
\bibitem{hawking} S. W. Hawking and I. G. Moss, Phys. Lett. {\bf
110B}, 35 (1982).
\bibitem{jst01} L. G. Jensen and P. H. Steinhardt, Nucl. Phys. {\bf B237}, 176 (1984); {\it ibid.} {\bf B317}, 693 (1989); J. Garriga and A. Vilenkin, Phys. Rev. D {\bf 57}, 2230 (1998); T. Banks,
arXiv:hep-th/0211160.
\bibitem{marvel01} U. Gen and M. Sasaki, Phys. Rev. D {\bf 61},
103508 (2000); K. Marvel and N. Turok, arXiv:0712.2719;
A. R. Brown and E. J. Weinberg, Phys. Rev. D {\bf
76}, 064003 (2007); E. J. Weinberg, Phys. Rev. Lett. {\bf 98},
251303 (2007); S.-H. Henry Tye, D. Wohns, and Y. Zhang,
Int. J. Mod. Phys. A {\bf 25}, 1019 (2010); K. Marvel and D. Wesley, J. High Energy Phys. 12 (2008) 034; S.-S. Xue, J. Korean Phys. Soc. {\bf 49}, 759 (2006); Int. J. Mod. Phys. A {\bf 24}, 3865 (2009).
\bibitem{bj05} T. Banks and M. Johnson, hep-th/0512141, A.
Aguirre, T. Banks, and M. Johnson, J. High Energy Phys. 08 (2006)
065; R. Bousso, B. Freivogel, and M. Lippert, Phys. Rev. D {\bf 74}, 046008 (2006).
\bibitem{kw04} K. Lee and E. J. Weinberg, Phys. Rev.
D {\bf 36}, 1088 (1987).
\bibitem{yms} Y. Kim, K. Maeda, and N. Sakai, Nucl. Phys.
{\bf B481}, 453 (1996); Y. Kim, S. J. Lee, K. Maeda, and N. Sakai,
Phys. Lett. B {\bf 452}, 214 (1999).
\bibitem{wl01} W. Lee, B.-H. Lee, C. H. Lee, and C. Park, Phys.
Rev. D {\bf 74}, 123520 (2006).
\bibitem{koh01} R.-G. Cai, B. Hu, and S. Koh, Phys. Lett. B {\bf 671}, 181 (2009).
\bibitem{landscape01} L. Susskind, hep-th/0302219; R. Bousso and
J. Polchinski, J. High Energy Phys. 06 (2000) 006; S. Kachru, R.
Kallosh, A. Linde, and S. P. Trivedi, Phys. Rev. D {\bf 68},
046005 (2003); B. Freivogel and L. Susskind, Phys. Rev. D {\bf
70}, 126007 (2004); D. I. Podolsky, J. Majumder, and N. Jokela, J.
Cosmol. Astropart. Phys. 05 (2008) 024; Q.-G. Huang and S.-H.
Henry Tye, Int. J. Mod. Phys. A {\bf 24}, 1925 (2009); D. Podolsky
and K. Enquvist, J. Cosmol. Astropart. Phys. 02 (2009) 007.
\bibitem{eternal01} A. Vilenkin, Phys. Rev. D {\bf 27}, 2848
(1983); A. D. Linde, Phys. Lett. B {\bf 175}, 395 (1986); A. H.
Guth, Phys. Rep. {\bf 333}, 555 (2000); A. Vilenkin,
gr-qc/0409055; D. I. Podolsky, Grav. Cosmol. {\bf 15}, 69 (2009).
\bibitem{nem} C. G. Callan, Jr. and S. Coleman, Phys. Rev. D {\bf 16}, 1762 (1977).
\bibitem{nemode} T. Tanaka and M. Sasaki, Prog. of Theor. Phys. {\bf 88}, 503 (1992); T. Tanaka, Nucl. Phys. {\bf B556}, 373 (1999); A. Khvedelidze, G. Lavrelashvili, and T. Tanaka, Phys. Rev. D {\bf 62}, 083501 (2000); G. Lavrelashvili, Nucl. Phys. Proc. Suppl. {\bf 88}, 75 (2000); G. Lavrelashvili, Phys. Rev. D {\bf 73}, 083513 (2006).
\bibitem{chw02} C. H. Lee and W. Lee, J. Korean Phys.
Soc. {\bf 45}, S1 (2004).
\bibitem{isr01} W. Israel, Nuovo Cimento B {\bf 44}, 1, (1966); {\it ibid.} B {\bf 48}, 463(E) (1967).
\bibitem{dyb} J. Hansen, D.-i. Hwang, and D.-h. Yeom, J. High Energy Phys. 11 (2009) 016; B.-H. Lee, W. Lee, and M. Minamitsuji, Phys. Lett. B {\bf 679}, 160 (2009); E. I. Guendelman and N. Sakai, Phys. Rev. D {\bf 77}, 125002 (2008); B.-H. Lee, C. H. Lee, W. Lee, S. Nam, and C. Park, Phys. Rev. D {\bf 77}, 063502 (2008); B.-H. Lee, W. Lee, S. Nam, and C. Park, Phys. Rev. D {\bf 75}, 103506 (2007).
\bibitem{ygh} J. W. York, Phys. Rev. Lett. {\bf 28}, 1082 (1972);
G. W. Gibbons and S. W. Hawking, Phys. Rev. D {\bf 15}, 2752
(1977).
\bibitem{prefactor} E. J. Weinberg, Phys. Rev. D {\bf 47}, 4614
(1993); G. V. Dunne and H. Min, Phys. Rev. D {\bf 72}, 125004
(2005); D. Metaxas, Phys. Rev. D {\bf 75}, 065023 (2007); {\it ibid.} D {\bf 78}, 063533 (2008).
\bibitem{hiscock} D. A. Samuel and W. A. Hiscock, Phys. Lett. B {\bf 261}, 251 (1991); Phys. Rev. D {\bf 44}, 3052 (1991).
\bibitem{gerlach} U. H. Gerlach, Phys. Rev. D{\bf 28}, 761 (1983).
\bibitem{dvil} A. Vilenkin, Phys. Lett. {\bf 133B}, 177 (1983).
\bibitem{dip} J. Ipser and P. Sikivie, Phys. Rev. D {\bf 30}, 712
(1984).
\end{thebibliography}
\end{document}